\documentclass[final]{article}
\usepackage{graphicx}
\usepackage{rotating}
\usepackage{amssymb}
\usepackage{mathptmx}
\usepackage[numbers]{natbib}
\usepackage{subscript}
\usepackage{gensymb} 
\usepackage{fancyhdr}
\usepackage{lastpage}
\usepackage{abstract}
\usepackage{hyphenat}
\usepackage{marvosym}
\usepackage{ragged2e}
\usepackage{subscript}
\usepackage{booktabs}
\usepackage{tabularx}
\usepackage{xcolor}
\usepackage[version=3]{mhchem}

\usepackage[vcentering,dvips,left=3.2cm,right=3.2cm,top=4cm,bottom=4cm]{geometry}

\usepackage{authblk}

\definecolor{AtomictAngerine}{rgb}{1.0, 0.6, 0.4}
\definecolor{UniBlue}{RGB}{83,121,170}

\usepackage{hyperref}
\hypersetup{
  breaklinks,
  pdfstartview=XYZ,
  bookmarks=true,
  colorlinks=true,
  linkcolor=UniBlue,
  urlcolor=UniBlue,
  citecolor=UniBlue,
  pdftex,
  bookmarks=true,
  linktocpage=true,   
  hyperindex=true
}

\usepackage{fancyhdr}
\renewenvironment{abstract}{%
  \hfill\par
    {\noindent\bfseries Abstract\par}
    \noindent\rule{\textwidth}{0pt}}
    {\par\noindent\rule{\textwidth}{0pt}
}

\makeatletter

\renewcommand\@maketitle{%
  \hfill
  \hspace*{-2em}
  \begin{minipage}{\textwidth}
  \vskip 2em
  \let\footnote\thanks 
  {\noindent\LARGE\bfseries\@title\par}
  \vskip 1.5em
  {\noindent\large\@author\par}
  \end{minipage}
  \vskip 1em \par
}
\makeatother

    {\par\addvspace{17pt}\small\rmfamily\noindent {\bfseries Acknowledgements}%
    \small\rmfamily\noindent}%
    {}%

\newcommand{\note}{\footnote}
\newcommand{\affiliation}{\affil}

\makeatletter

\setcitestyle{square}

\begin{document}

\title{\nohyphens{Measuring the electron neutrino mass with improved sensitivity: the HOLMES experiment}}

\author[a]{A.~Giachero\note{Corresponding author \Telefon~+39 02-6448-2456, \Letter~Andrea.Giachero@mib.infn.it}}
\author[b]{B.K.~Alpert}
\author[b]{D.T.~Becker}
\author[b]{D.A.~Bennett}
\author[e,d]{M.~Biasotti}
\author[f,a]{C.~Brofferio}
\author[d]{V.~Ceriale}
\author[a]{G.~Ceruti}
\author[e,d]{D.~Corsini}
\author[g]{P.K.~Day}
\author[e,d]{M.~De~Gerone}
\author[h]{R.~Dressler}
\author[f,a]{M.~Faverzani}
\author[a]{E.~Ferri}
\author[b]{J.W.~Fowler}
\author[d]{E.~Fumagalli}
\author[d]{G.~Gallucci}
\author[b]{J.D.~Gard}
\author[e,d]{F.~Gatti}
\author[a,b]{J.P.~Hays-Wehle}
\author[h]{S.~Heinitz}
\author[b]{G.C.~Hilton}
\author[i]{U.~K\"oster}
\author[j,k]{M.~Lusignoli}
\author[b]{J.A.B.~Mates}
\author[c]{S.~Nisi}
\author[f,a]{A.~Nucciotti}
\author[d]{A.~Orlando}
\author[d]{L.~Parodi}
\author[a]{G.~Pessina}
\author[e,d]{G.~Pizzigoni}
\author[f,a]{A.~Puiu}
\author[f,a]{S.~Ragazzi\note{currently at INFN - Laboratori Nazionali del Gran Sasso (LNGS), Assergi (L'Aquila) I-67010, Italy }}
\author[b]{C.D.~Reintsema}
\author[l]{M.~Ribeiro Gomes}
\author[b]{D.R.~Schmidt}
\author[h]{D.~Schumann}
\author[d]{F.~Siccardi}
\author[f,a]{M.~Sisti}
\author[b]{D.S.~Swetz}
\author[f,a]{F.~Terranova}
\author[b]{J.N.~Ullom}
\author[b]{L.R~Vale}

\affiliation[a]{INFN - Sezione di Milano Bicocca, Milano I-20126 - Italy}
\affiliation[b]{National Institute of Standards and Technology (NIST), Boulder, Colorado 80305, USA}
\affiliation[c]{INFN - Laboratori Nazionali del Gran Sasso (LNGS), Assergi (L'Aquila) I-67010, Italy}
\affiliation[d]{INFN - Sezione di Genova, Genova I-16146, Italy}
\affiliation[e]{Dipartimento di Fisica, Universit\`{a} di Genova, Genova I-16146, Italy}
\affiliation[f]{Dipartimento di Fisica, Universit\`{a} di Milano-Bicocca, Milano I-20126, Italy}
\affiliation[g]{Jet Propulsion Laboratory (JPL), Pasadena, California 91107, USA}
\affiliation[h]{Paul Scherrer Institut (PSI), 5232 Villigen, Switzerland}
\affiliation[i]{Institut Laue-Langevin (ILL), 38000 Grenoble, France}
\affiliation[j]{Dipartimento di Fisica, Sapienza Universit\`{a} di Roma, Roma I-00185, Italy}
\affiliation[k]{INFN - Sezione di Roma, Roma I-00185, Italy}
\affiliation[l]{CENTRA - Multidisciplinary Centre for Astrophysics, University of Lisbon, 1049-001 Lisbon, Portugal}


\maketitle

\begin{abstract}
HOLMES is a new experiment aiming at directly measuring the neutrino mass with a sensitivity below 2\,eV. HOLMES will perform a calorimetric measurement of the energy released in the decay of \textsuperscript{163}Ho. The calorimetric measurement eliminates systematic uncertainties arising from the use of external beta sources, as in experiments with spectrometers. This measurement was proposed in 1982 by A. De Rujula and M. Lusignoli, but only recently the detector technological progress has allowed to design a sensitive experiment. HOLMES will deploy a 1000 pixels array of low temperature microcalorimeters with implanted \textsuperscript{163}Ho nuclei. HOLMES, besides being an important step forward in the direct neutrino mass measurement with a calorimetric approach, will also establish the potential of this approach to extend the sensitivity down to 0.1\,eV and lower.


The detectors used for the HOLMES experiment will be Mo/Cu bilayers TESs (Transition Edge Sensors) on SiN\textsubscript{x} membrane with gold absorbers. Microwave multiplexed rf-SQUIDs are the best available technique to read out large array of such detectors. An extensive R\&D activity is in progress in order to maximize the multiplexing factor while preserving the performances of the individual detectors. To embed the \textsuperscript{163}Ho into the gold absorbers a custom mass separator ion implanter is being developed.

The current activities are focused on the the single detector performances optimization and on the \textsuperscript{163}Ho isotope production and embedding. A preliminary measurement of a sub-array of $4\times 16$ detectors is planned late in 2017. In this contribution we present the HOLMES project with its technical challenges, its status and perspectives.
\end{abstract}

\justify\section{Introduction}
Experiments on oscillations demonstrated that neutrinos are massive particles, but, since they are sensitive only on the difference between the square of the neutrino mass eigenvalues, they cannot provide information about the absolute scale~\cite{Capozzi2016}. The study of the energy distribution at the beta, or electron capture (EC), decay spectrum end-point is currently the one and only experimental method which can provide a model independent measurement of the absolute scale of neutrino mass~\cite{Drexlin}. Isotopes with low decay energy $Q$ are preferable in order to maximize the statistics in the energy region of the end-point.  

\begin{figure}[!t]
\centering 
\includegraphics[width=\textwidth,clip]{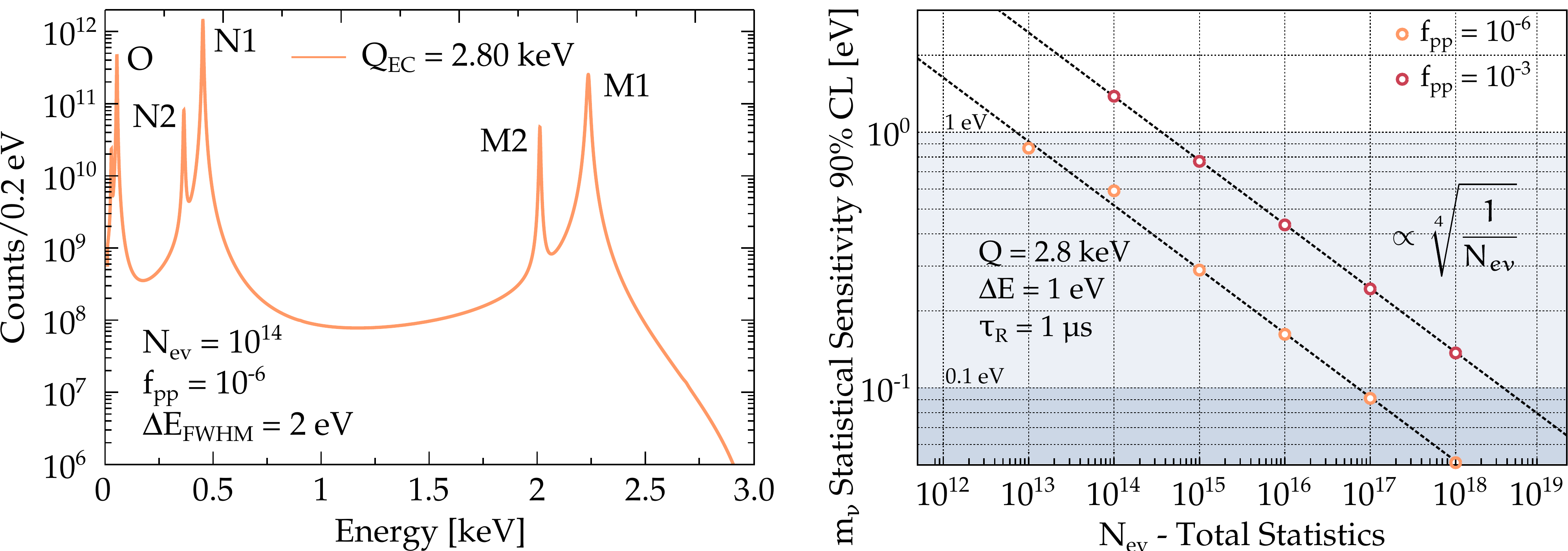}
\caption{\label{fig:163Hospectrum} Left: $^{163}$Ho total absorption spectrum. Right: $^{163}$Ho decay experiments statistical sensitivity dependence on the total statistics for two different values of pile-up fraction~\cite{Nucciotti_Sensitivity}.}
\end{figure}

An interesting approach suitable for a direct determination of the neutrino mass, is the calorimetric measurement of the energy released by the $^{163}$Ho decay, as proposed by De Rujula and Lusignoli in 1982~\cite{DeRujula_1}. $^{163}$Ho decays via electron capture on $^{163}$Dy with a very low $Q_{EC}$-value of 2.883\,keV~\cite{PENNING}: the proximity of the M1 resonance to the end-point (figure \ref{fig:163Hospectrum}, left) further enhances the statistics in the region of interest (ROI), making $^{163}$Ho a very promising choice for the neutrino mass determination. In a calorimetric measurement, the energy released in the decay process is entirely contained into the detector, except for the fraction taken away by the neutrino. This approach eliminates both the problematics connected to the use of an external source and the systematic uncertainties arising from decays on excited final states. The most suitable detectors for this type of measurement are low temperature thermal detectors, where all the energy released into an absorber is converted into a temperature increase that can be measured by a sensitive thermometer directly coupled to the absorber~\cite{Nucciotti}.

Applying a frequentist Monte Carlo approach~\cite{Nucciotti_Sensitivity} it is possible to demonstrate that for a $^{163}$Ho calorimetric experiment the statistical sensitivity on the neutrino mass is proportional to $\sqrt[4]{1/N_{ev}}$, where $N_{ev}$ is the number of the detected event. This numerical approach allows to include the detector energy ($\Delta E_{FWHM}$) and time ($\tau_r$) resolutions and the pile-up fraction ($f_{pp}=\tau_r\cdot A_{EC}$, where $A_{EC}$ is the activity per pixel). 

Considering a high energy and time resolution ($\simeq\,1$\,eV and $\simeq\,1$\,$\mu$s, respectively) and a pile-up fraction within the range $f_{pp}=10^{-3}-10^{-6}$, a sub-eV neutrino sensitivity can be reached collecting a total number of events around $N_{ev}> 10^{13} -  10^{15}$ (figure \ref{fig:163Hospectrum}, right). To further lower the sensitivity below 0.1 eV a total number of event around $N_{ev}> 10^{17} -  10^{19}$ are needed. Considering two different pixel activities of 1\,Bq and 1000\,Bq this means a total number of collected event around $2\cdot 10^{9}$\,det$\cdot$years and $10^{8}$\,det$\cdot$years respectively. To achieve these requirements a megapixel arrays experiment is needed where the multiplexing factor and the read-out bandwidth play a crucial role for future developments. 

Only in recent times, low temperature microcalorimeters reached the necessary maturity to be used in a large scale experiment with good energy and time resolutions and are therefore the detectors of choice for a sub-eV holmium experiments. Within this framework, the HOLMES experiment~\cite{HOLMES} will perform a calorimetric measurement of the energy released in the electron capture decay of $^{163}$Ho. The main goal of HOLMES is to prove the potential and the scalability of this technique for a future megapixel experiment. Other, but not secondary goals, are to reach a neutrino mass statistical sensitivity below 2\,eV, assess the calorimetric EC Q-value and the systematic errors specific of a long calorimetric measurement. In order to reach this sensitivity HOLMES will collect about $3 \cdot 10^{13}$ events by deploying an array of 1024 detectors, composed by 16 sub-arrays, $4\times 16$ pixels each. The total amount of implanted $^{163}$Ho nuclei will be about $6.5\cdot 10^{16}$, equivalent to 18\,$\mu$g for a total activity of 300\,dec/s/pixel. The  baseline for HOLMES are Mo/Cu TESs (Transition Edge Sensors) on SiN$_x$ membrane with  $^{163}$Ho-implanted gold absorbers, read out with rf-SQUIDs, for microwave multiplexing purposes. The target for the instrumental energy resolution and time resolutions are around $3-5$\,eV FWHM and $10\,\mu$s respectively.

A preliminary measurement of a sub-array of 64 detectors, identical to the final ones, is planned by the end of 2017. The goal of this measurement is to validate all the production procedures defined and all detector parts designed for HOLMES. Obtained results will also provide precise information on the atomic and nuclear parameter of $^{163}$Ho electron capture decay, complementing the data provided by ECHo~\cite{ECHo} and NuMECS~\cite{NuMECS2} experiments. The final measurement with the full $64\times 16$ array is expected to start in late 2018. HOLMES is characterized by the following main tasks: $^{163}$Ho isotope production, $^{163}$Ho source embedding, single detector optimization and array engineering, multiplexed read-out and data-analysis. We outline in this work these tasks and their current status after two years of developments.

\section{$^{163}$Ho isotope production}
The $^{163}$Ho isotope is not present in nature. One of the methods for production is neutron irradiation of Erbium enriched in $^{162}$Er. The reaction \ce{^{162}Er(n{,}\gamma)^{163}Er}  and the subsequent EC decay of $^{163}$Er with a short half-life of 75 min leads to $^{163}$Ho. HOLMES uses the high-flux reactor of ILL (Institut Laue Langevin, Grenoble, France) with a thermal neutron flux of $1.3\cdot 10^{15}$\,n/s/cm$^2$ at 90\% reactor power~\cite{ILL}. 

This high flux combined with the appreciable thermal cross-section of $\approx$ 19\,barn assures an efficient way to produce $^{163}$Ho: about $8\cdot 10^{15}$ atoms of $^{163}$Ho are expected to be produced per day per mg of $^{162}$Er. Provided self-shielding by $^{167}$Er and product burn-up of $^{163}$Ho are negligible. The chemical purification of the Er$_2$O$_3$ powder before irradiation and a chemical separation of holmium in Hot-cells after the neutron irradiation is performed at the Laboratory of Radiochemistry (LRC) at the Paul Scherrer Institute (PSI, Villigen, Switzerland). 

Two samples of enriched  Er$_2$O$_3$ have been irradiated at ILL and processed at PSI. Inductive Coupled Plasma Mass Spectroscopy (ICPMS) analysis performed at LNGS (Gran Sasso Underground Laboratory, L'Aquila, Italy) and PSI on the two samples before and after chemical purification at PSI demonstrated a production of a sufficiently radiopure sample of about 35\,MBq of $^{163}$Ho. The global efficiency of the chemical purification at PSI has been assessed to be about 79\%. For the HOLMES research program the total estimated amount of $^{163}$Ho required is about 220\,MBq. This translates into an irradiation at ILL for 50\,days of about 1100\,mg of Er$_2$O$_3$ enriched in $^{162}$Er at about 25\%. 

The chemical state of holmium influences the end point of the EC spectrum. In order to avoid this effect, Ho is embedded in the absorber in its metallic form. The reduction and distillation of pure metallic holmium is possible by heating the mixture of yttrium and holmium oxide above the melting point~\cite{Pizzigoni}. A Tantalum heated Knudsen cell is used for the reduction and distillation of holmium from the Er$_2$O$_3$. Holmium is thermally reduced at about 2000\,K using the reaction Ho$_2$O$_3$+2Y({\it met})$\rightarrow$2Ho({\it met})+ Y$_2$O$_3$. A new thermally optimized cell has been designed and installed in a new evaporation system set-up. The system is successfully running since February and it is being optimized. Tests are in progress with natural holmium oxide to tune the reduction process and assess the overall efficiency.

A sample of about 550\,mg of Er$_2$O$_3$ at 25.1\% has been purchased in July 2016. The irradiation of this sample and of about 100\,mg of  Er$_2$O$_3$ at 26.9\%, which have been recycled at PSI from the previously irradiated samples, will be performed from January to March 2017. Thus, during 2017 a total inventory of about 165\,MBq of $^{163}$Ho will be available for implantation.

\begin{figure}[!t]
\centering 
\includegraphics[width=0.7\textwidth,clip]{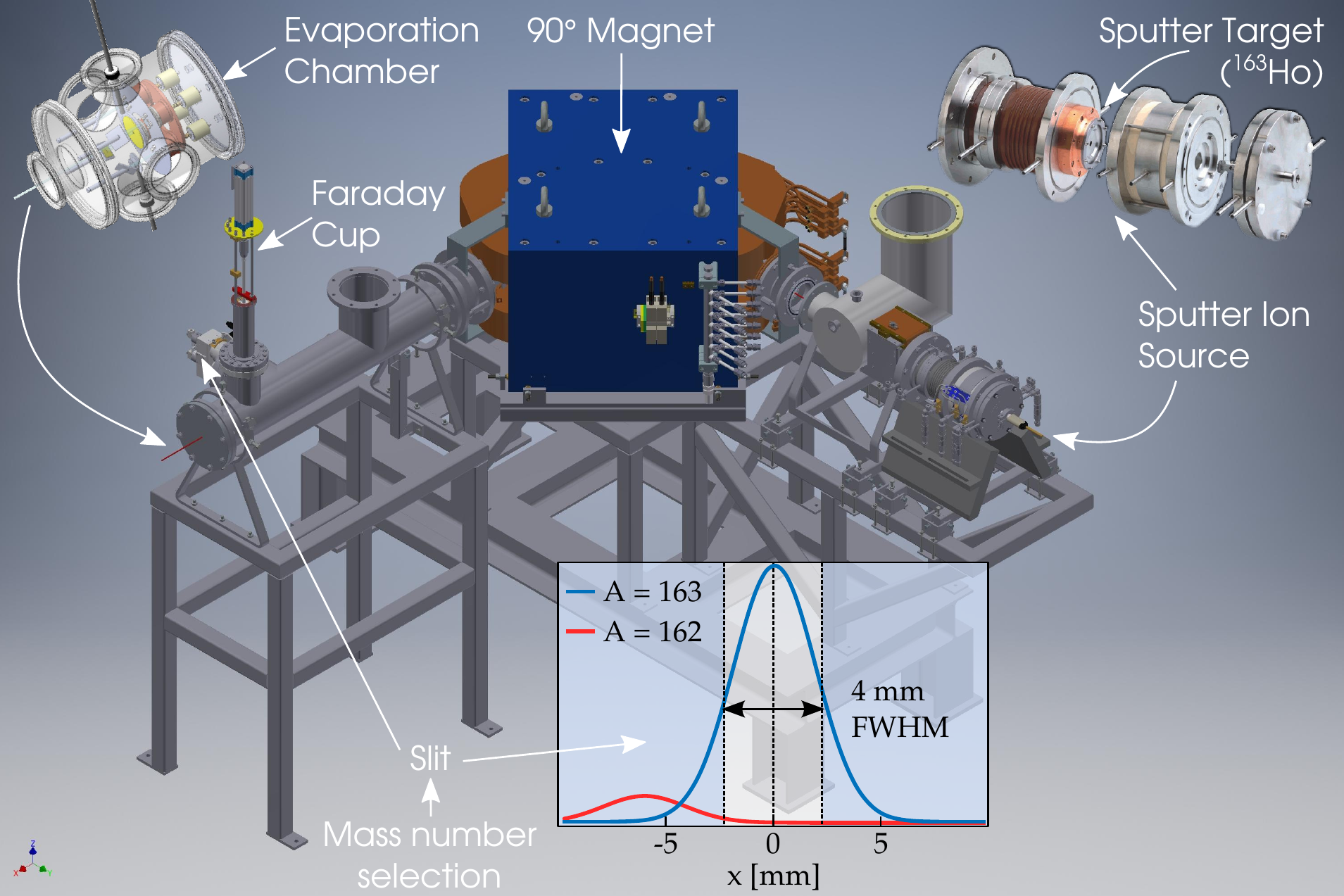}
\caption{\label{fig:implanter} The HOLMES custom ion implanter: from right to left, the Penning sputter ion source, the steering magnet, the mass analyzing magnet, and the evaporation chamber.}
\end{figure}

\section{$^{163}$Ho implanting system}
The $^{163}$Ho nuclei will be implanted into the microcalorimeter gold absorber by means of a custom ion implanter consisting of a Penning sputter ion source, a steering magnet, and a mass analyzing magnet (figure \ref{fig:implanter}). The latter parts of the system are designed to isolate $^{163}$Ho by means of mass separation, thereby eliminating other trace contaminants not removed by chemical methods at PSI. One such contaminant is  $^{166m}$Ho, which could be produced during neutron irradiation. $^{166m}$Ho is a $\beta$ decaying isotope with a half life of about 1132\,years~\cite{Nedjadi_166Ho}, and decay events below 5\,keV, close to the region of interest. Therefore, removal of this isotope is potentially critical for avoiding extra background.

The metallic cathode for the ion source will be made out of a metallic holmium pellet containing the $^{163}$Ho produced in the reduction and distillation process. The implanter is integrated with a vacuum chamber (the target chamber) which also allows a simultaneous gold evaporation to control the $^{163}$Ho concentration and to deposit the final gold layer to complete the $^{163}$Ho embedding. Simulations are in progress to design a beam focusing stage with the purpose of decreasing the beam size on the target, thereby improving the geometrical efficiency of the implantation process.

The setting up of the laboratory in Genova where the embedding system will be hosted, is in progress. The ion source and the magnet have been delivered by the end of 2016. The system commissioning will start in 2017. Preliminary tests will be performed by using stable holmium or $^{166m}$Ho. This initial phase will allow to assess the efficiency of the entire process, from the neutron irradiated Er$_2$O$_3$ powder to the detector absorber embedding. The target chamber is being set-up in Milano-Bicocca and it will be initially used to tune the detector gold absorber fabrication process without holmium implantation. It will be finally integrated with the implanter.

\begin{figure}[!t]
\centering 
\includegraphics[width=\textwidth,clip]{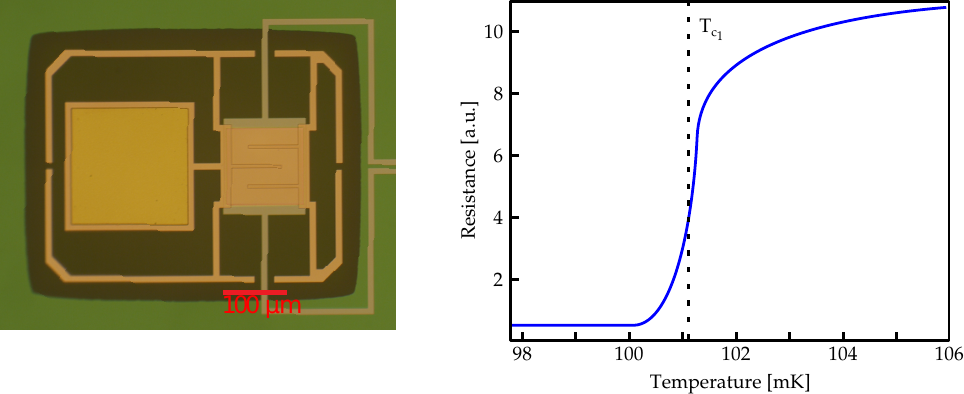}
\caption{\label{fig:TES} Left: picture of the developed Mo/Cu TES with Au absorber placed alongside (\textit{side-car} design). Right: TES resistance as a function of the temperature. The critical temperature $T_c$ is around 100\,mK}
\end{figure}

\section{TES sensors optimization}
HOLMES will use TES based microcalorimeters adapted from use in high-resolution soft X-ray spectroscopy to this specific application. The single pixel will be Molybdenum-Copper bilayer TES on SiN membrane with 2\,$\mu$m-thick gold absorbers, which ensures a 99.99998\% (99.9277\%) probability of stopping the electrons (photons) coming from the decay of $^{163}$Ho. Development of the single TES pixel design may continue, but a suitable protype has already been identified and proven.

The requirement of low pile-up probability sets strict constraints on the detector response. The decay time is set by the ratio between the thermal capacity $C$ of the absorber and the thermal conductance $G$ toward the bath. Since the heat capacity is constrained by requiring the full containment of the energy in the absorber, the only way to increase the speed is to increase $G$. A series of different TES prototypes suspended on a SiN membrane were produced and tested at the National Institute for Standard and Technology (NIST, Boulder, Co, USA). 

The selected design is a $125 \times 125\,\mu$m$^2$ Mo/Cu TES with three normal metal bars built on a SiN membrane (figure \ref{fig:TES}, left). A $200 \times 200\,\mu$m$^2$ gold absorber is placed alongside the TES to avoid proximity effect between the gold and the sensor itself (\textit{side-car} design). The thermal conductance $G$ is increased by the addition of a thermal radiating perimeter that increases the conductance in this 2-d geometry. This thermalizing perimeter increases the thermal conductance $G$ without raising the heat capacity above 0.8\,pJ/K. With this and other techniques, any $G$ from 40\,pW/K up to 1\,nW/K is achievable~\cite{Hays-Wehle}. Devices with increased $G$ have faster decay times, as predicted. The critical temperature, $T_c$ resulted around 100\,mK (figure \ref{fig:TES}, right), as designed, and to maximize the performances the detectors have to work at a base temperature around 60\,mK. Pixels engineered to have a decay time around 150\,$\mu$s showed sufficient resolution for the intended application: few eVs at $Q_{EC}$. By using the TES thermal model developed at NIST~\cite{Irwin2005}, along with signal processing simulation software, it is possible to predict that the effective time resolution achievable with such detectors will be limited by the sampling frequency of the read-out system. Therefore, although a rise time as fast as a few $\mu$s has been obtained, a rise time to around 10\,$\mu$s was preferred in order to maximize the multiplexing factor without affecting the energy resolution. 

The final HOLMES TESs detector arrays will be fabricated in a two step process. The devices will be provided by NIST with a 1\,$\mu$m gold layer and they will be further processed in Italy. The first process will be the deposition in the target chamber by means of simultaneous ion implantation and ion beam sputtering of a thin (few 100\,\AA{}) layer of Au/$^{163}$Ho, then the gold absorber will be completed in the target chamber by sputtering a second 1\,$\mu$m gold layer to fully encapsulate the $^{163}$Ho source. Finally the fabrication will be completed by fully releasing the Si$_2$N$_3$  membranes by means of a Deep Reactive Ion Etching which will be performed at the Istituto di Fotonica e Nanotecnologie - CNR in Roma. Design and simulation work along with preliminary tests on dummy samples from NIST are in progress to define the details of this two-step fabrication process.

\begin{figure}[!t]
\centering 
\includegraphics[width=\textwidth,clip]{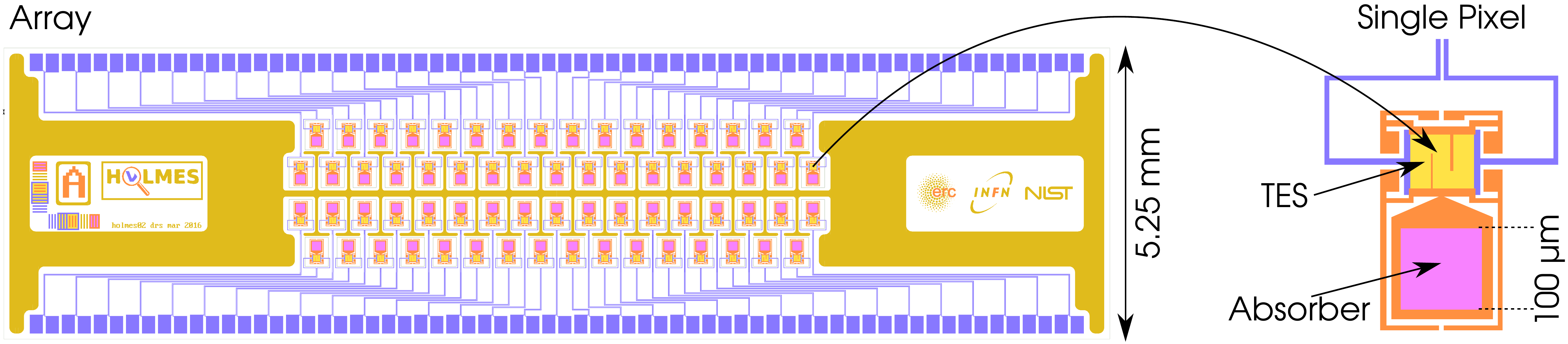}
\caption{\label{fig:64array} Design of the HOLMES $4\times 16$ sub-array (left) and of the single pixel in the \textit{side-car} form (right).}
\end{figure}

The design of the $4\times16$ sub-arrays which will compose the HOLMES array is finalized (figure \ref{fig:64array}). The design aims to 1) minimize the signal bandwidth limitations due to stray self-inductance of the read-out leads,  2) minimize the signal cross-talk due to mutual inductance between read-out lines, and 3) maximize the geometrical filling for an optimal implantation efficiency. The first sub-arrays will be fabricated at NIST a the beginning of 2017. They will be used for testing the two-step fabrication process, the ion implantation, and to assess the thermal effect of high holmium concentrations in the absorbers. 

\section{Multiplexing read-out and TESs characterization}
The 1024 channels of the HOLMES array will be read out with the microwave multiplexing system ($\mu$Mux)~\cite{uMux} developed in collaboration with NIST. 

This scheme uses dissipation-less radio-frequency (rf) SQUIDs to transduce TES current into a frequency shift of a resonator. In addiction a flux-ramp modulation is applied to the SQUID  to linearize the response~\cite{ramp}. The modulated signals are read out coupling the rf-SQUID to superconducting lambda/4-wave resonators in the GHz range and using the homodyne detection technique. Tuning the resonators at different frequencies it is straightforward to multiplex many RF carriers. Microwave multiplexing is the most suitable system for HOLMES, since it provides a larger bandwidth for the same multiplexing factor (number of multiplexed detectors). This novel multiplexing and read-out approach was demonstrated for the first time for gamma-ray spectroscopy~\cite{Noroozian} and has been proposed for many current and future applications based on superconducting transition-edge sensor where fast pulse response is required.
 


The $\mu$Mux is suitable for a fully digital approach based on the Software Defined Radio (SDR) technique. In this scenario, a comb of frequency carriers are generated by digital synthesis in the MHz (base-band) range and up-converted to the GHz range (RF band) by IQ-mixing. The GHz comb is sent to the cold $\mu$Mux chips coupled to the detectors through one semi-rigid cryogenic coax cable, then it is amplified by a low temperature and low noise High Electron Mobility Transistor (HEMT), and finally it is sent back to room temperature through another coax cable. The output signal is down-converted by IQ-mixing and the \textit{channelization} (i.e. individual channel signal recovery) is performed by software on the digitized signal. A tiny change in temperature due to a X-ray interaction results in a change in the input flux to the rf-SQUID that causes a change of resonant frequency of the $\mu$Mux resonator and hence a variation of the phase of the transmitted signal. By monitoring the resonances with the technique explained above it is possible to recover the changes in the resonance frequency and phase of each resonator.

\begin{figure}[!t]
\centering 
\includegraphics[width=\textwidth,clip]{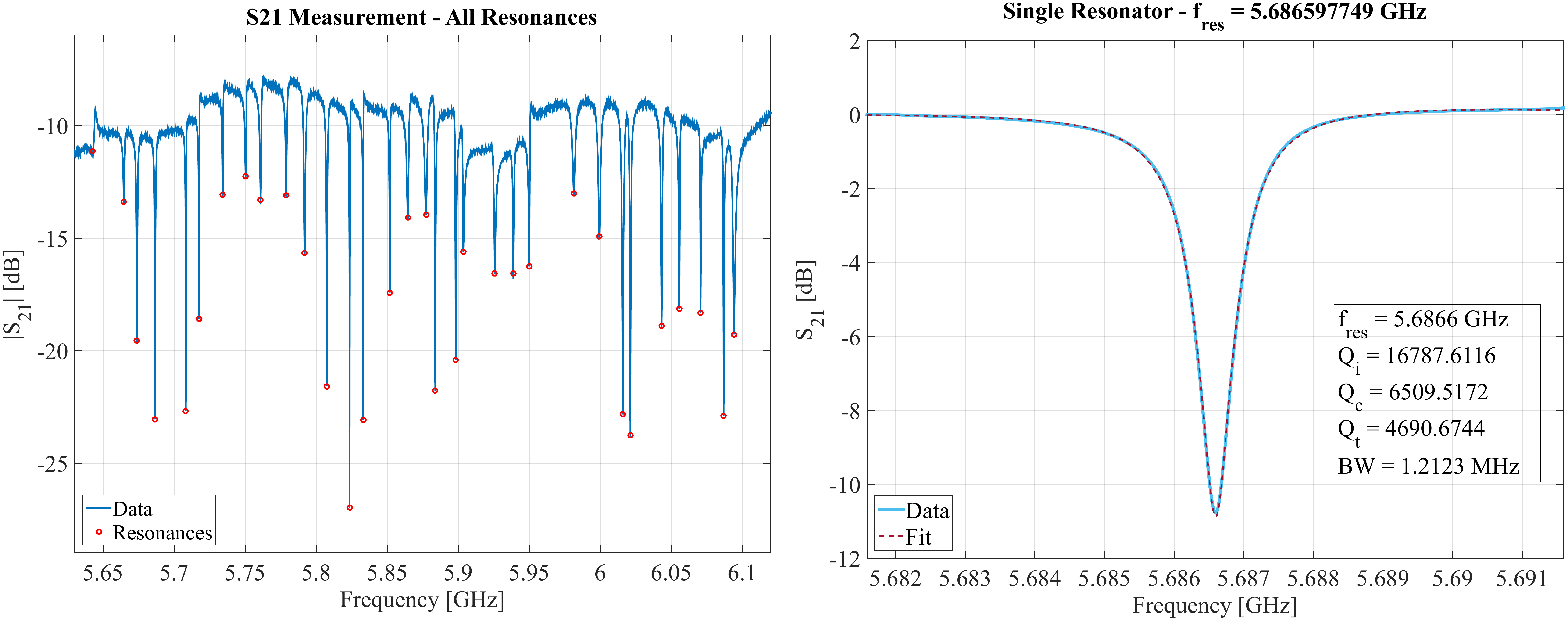}
\caption{\label{fig:mux} Characterization of the $\mu$MUX16a developed for HOLMES. Left: wide-span frequency measurement of the entire chip resonances. Right: example of resonance fit used to extrapolate the resonator parameters.}
\end{figure}

The first prototype of chip multiplexers developed for HOLMES ($\mu$MUX16a) consists of 33 quarter-wave coplanar waveguide (CPW) microwave resonators made from a 200\,nm thick Nb film deposited on high-resistivity silicon ($\rho >10\,\mbox{k}\Omega\cdot$cm). This chip is designed to have resonances with bandwidth of 1 to 2\,MHz for reading out demodulated signals with a sampling frequency up to 500\,kS/s. Chips were developed at NIST and characterized at NIST and in Milano.
Results showed 31/33 resonators usable for microwave readout, two resonators collided to one and one has an irregular shape (figure \ref{fig:mux}). The spacing between adjacent resonances resulted 14\,MHz, value that assure a negligible cross-talk. The resonator average bandwidth resulted 1.6\,MHz while the average SQUID noise resulted 4$\mu\Phi_0/\sqrt{\mbox{Hz}}$, not critical for the energy resolution. These results demonstrated that the chips $\mu$MUX16a is fully compliant with the HOLMES specifications. A new design for increasing the resonators uniformity is in progress.

The $\mu$MUX16a has also been tested coupled with the first $4\times 6$ TES array prototype developed at NIST. The array accommodates different perimeter/absorber configurations in order to study the detector response. An interface chip is used to provide a bias shunt resistor of $R_{\scalebox{0.7}{\mbox{shunt}}} = 0.33$\,m$\Omega$ in parallel with each TES and a wirebond-selectable Nyquist inductor, $L_{N}$, placed in series to tune the pulse rise-times. The $\mu$MUX chip, TESs and interface chips were placed inside a copper holder connectorized with two SMAs for the feedline and a Micro-D for the bias and ramp lines. A X-ray source ($^{55}$Fe) was mounted externally to the holder and it irradiated the sensitive area of the detectors through a silicon collimator. 

\begin{figure}[!t]
\centering 
\includegraphics[width=\textwidth,clip]{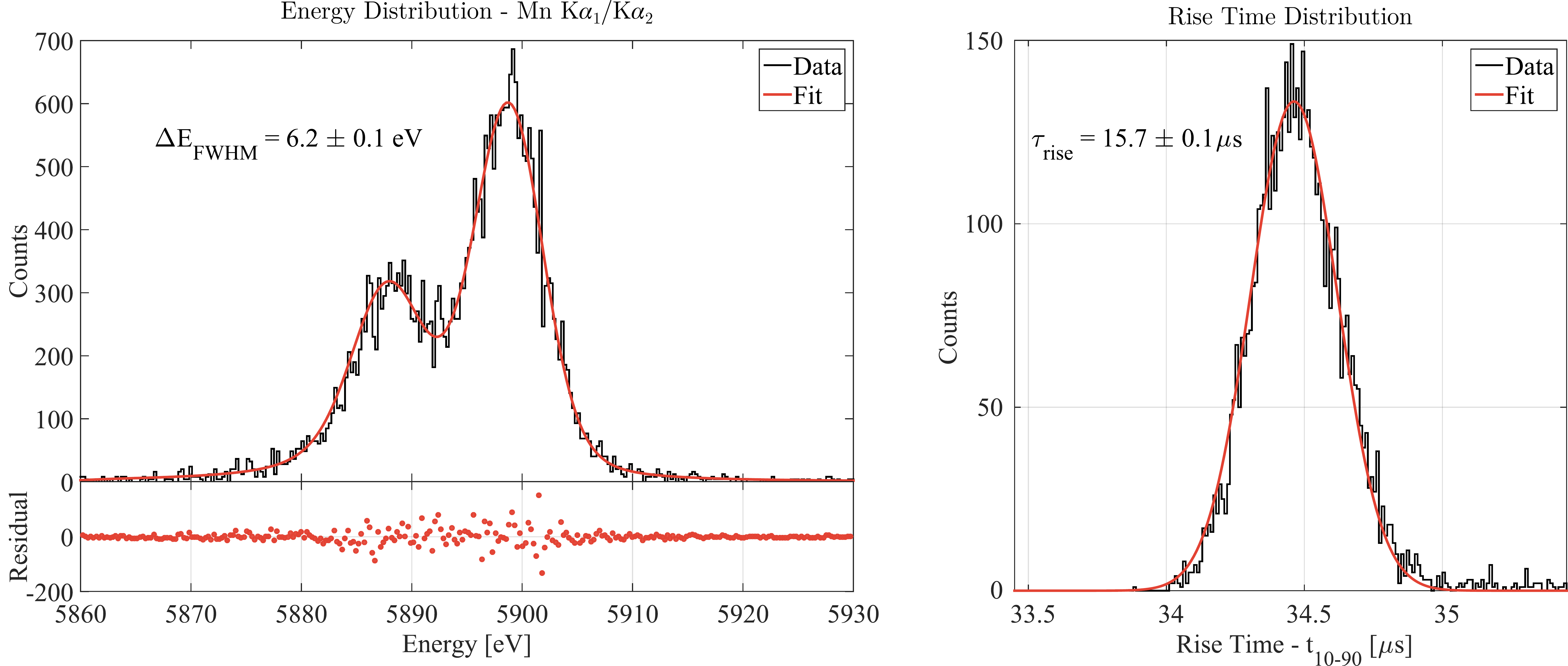}
\caption{\label{fig:res} Obtained energy (left) and time (right) resolutions for one class of the produced TESs by using a $^{55}$Fe X-ray source.}
\end{figure}

Tests were performed by using a two-channel system developed in Milano and based on commercial components, that exploits homodyne technique. This system is an improved version of the one presented in~\cite{Giachero} with a faster ADC board ($f_s = 250$\,MS/s) that allows to acquire fast detectors thanks to a ramp demodulation rate of $f_{\scalebox{0.7}{\mbox{ramp}}} =400-500$\,kHz, which is the effective demodulated signal sampling rate. The measured TESs were previously characterized at NIST with standard techniques. When operated at a bias point that is 20\% of the normal state resistance $R_n$ ($9\,\mbox{m}\Omega$) they showed a sensitivity to temperature $\alpha$ of 60, a current sensitivity $\beta$ of 1.8, and a noise parameter $M$ of 1.5. The $C$ and $G$ thermal parameters resulted around 0.8\,pJ/K and 600\,pW/K, respectively. Results obtained with the two-channel system developed in Milano showed a $^{55}$Mn K$\alpha_1$/K$\alpha_2$ peak separation with an energy resolution of about 6\,eV (figure \ref{fig:res}, left) and a noise level of about 5\,eV. The same detector at NIST was measured with a more standard Time Domain Multiplexer and showed an energy resolution of 4 eV. Work is in progress to further improve the cryogenic set-up and match the TDM results.

The fast data acquisition system combined with the large-bandwidth resonators allowed to read out TES signals with 16\,$\mu$s of exponential rise time constant ($\simeq 35$\,$\mu$s 10\% to 90\%) (figure \ref{fig:res}, right). This value is higher than the one needed for HOLMES but it was limited only by the Nyquist inductor $L_{N}$, in series to the TES, and not by the TES itself. Preliminary results from measurements currently in progress show that reducing the $L_{N}$ from 64\,nH (8 turns) to 50\,nH (6 turns) a rise time constant of 9\,$\mu$s ($\simeq 20$\,$\mu$s 10\% to 90\%) is achievable, fulfilling the HOLMES requirement.

In its final configuration HOLMES will realize a SDR multiplexed read-out exploiting the Reconfigurable Open Architecture Computing Hardware (ROACH2) board with a Xilinx Virtex6 FPGA. The complete system is composed of a digital signal processing board (ROACH2), a DAC (for comb generation) and ADC (512\,MS/s, 12\,bit, 2 channels) boards, an IF board (for signal up- and down-conversion), and SFP+ GbE interfaces optically decoupled for fast data transfer. Software, firmware and set-up are developed in collaboration with NIST. Tests with a preliminary version of the firmware for the multiplexing of 4 channels showed encouraging results. An expanded version for 32 channels is in development and it will be ready in 2017. To read out the full 1024 pixel array a total of 32 ROACH2-based systems are required.

\section{Conclusion}
The measurement of the end point of nuclear beta or electron capture (EC) decays spectra is the only model-independent experimental tool for accessing the absolute neutrino mass. Experiments in the near future will achieve sub-eV neutrino mass sensitivity. The HOLMES experiment will perform a direct measurement of the neutrino mass by using low temperature microcalorimenter with $^{163}$Ho-implanted absorber.

The produced microwave multiplexer and TES pixel prototypes fulfill the HOLMES requirements, and will be finalized in the final configuration in 2017. A sample of 25.1\% enriched Er$_2$O$_3$ will be irradiated at ILL, consenting to achieve a total inventory of about 165\,MBq of $^{163}$Ho, which will be ready in mid-2017; this amount ensures about one half of the total required. Tests to tune the holmium reduction process and assess the overall efficiency  are in progress. The implanter design is finalized and delivery is scheduled before the end of 2016. The commissioning will start in 2017. Starting from 2017 the first $4 \times 16$ linear arrays will be produced and characterized, and implanting tests will start after the implanter commissioning. In parallel the 32-channel read-out and multiplexing system will be finalized. At the end of 2017 one month measurement with one $4 \times 16$ sub array is scheduled. 2018 will be dedicated at the $32 \time 32$ array development and commissioning.

\section*{Acknowledgements}
This work was supported by the European Research Council (FP7/2007-2013) under Grant Agreement HOLMES no. 340321. We also acknowledge the support from INFN through the MARE project and from the NIST Innovations in Measurement Science program for the TES detector development.

\bibliographystyle{JHEP}
\bibliography{holmes-IPRD16-proceedings}

\end{document}